\documentclass[prb,twocolumn,showpacs,superscriptaddress, amsmath,amssymb]{revtex4}

\usepackage{graphicx}
\usepackage{dcolumn}
\usepackage{bm}
\usepackage[dvips]{color}
\usepackage{color}

\newcommand{\vect}[1]{\mathbf{#1}}

\begin{document}
\preprint{APS/123-QED}

\title{Magnetic structure and domain conversion of quasi-2D frustrated antiferromagnet CuCrO$_2$ probed by NMR.}

\author{Yu. A. Sakhratov}
\affiliation{National High Magnetic Field Laboratory, Tallahassee, Florida
32310, USA} \affiliation{Kazan State Power Engineering University, 420066
Kazan, Russia}

\author{L. E. Svistov}
\email{svistov@kapitza.ras.ru}
 \affiliation{P. L. Kapitza Institute for
Physical Problems RAS, 119334 Moscow, Russia}

\author{P. L. Kuhns}
\affiliation{National High Magnetic Field Laboratory, Tallahassee, Florida
32310, USA}

\author{H. D. Zhou}
\affiliation{National High Magnetic Field Laboratory, Tallahassee, Florida
32310, USA} \affiliation{Department of Physics and Astronomy, University of
Tennessee, Knoxville, Tennessee 37996, USA}

\author{A. P. Reyes}
\affiliation{National High Magnetic Field Laboratory, Tallahassee, Florida
32310, USA}

\date{\today}

\begin{abstract}
We have carried out $^{63,65}$Cu NMR spectra measurements in magnetic field up to about 15.5~T on single crystal of a multiferroic triangular-lattice antiferromagnet CuCrO$_2$. The measurements were performed for perpendicular and parallel orientation of the magnetic field with respect to the $c$-axis of the crystal, and the detailed angle dependence of the spectra on the magnetic field direction within $ab$-plane was studied. The shape of the spectra can be well described in the model of spiral spin structure proposed by recent neutron diffraction experiments. When field is rotated perpendicular to crystal $c$-axis, we observed, directly for the first time, a remarkable reorientation of spin plane simultaneous with rotation of the incommensurate wavevector by quantitatively deducing the conversion of less energetically favorable domain to a more favorable one. At high enough fields parallel to $c$-axis, the data are consistent with either a field-induced commensurate spiral magnetic structure or an incommensurate spiral magnetic structure with a disorder in the $c$ direction, suggesting that high fields may have influence on interplanar ordering.
\end{abstract}


\maketitle

\section{Introduction}

The problem of an antiferromagnet on a triangular planar lattice has been
intensively studied
theoretically.~\cite{Kawamura_1985,Korshunov_1986,Anderson_1987,Plumer_1990,Chubukov_1991}
The ground state in the Heisenberg and XY models is a ``triangular'' planar
spin structure with three magnetic sublattices arranged 120$^\circ$ apart. The
orientation of the spin plane is not fixed in the exchange approximation in the
Heisenberg model. The applied static field does not remove the degeneracy of
the classical spin configurations. Therefore the usual small corrections such
as quantum and thermal fluctuations, and relativistic interactions in the
geometrically frustrated magnets play an important role in the formation of the
equilibrium state.~\cite{Korshunov_1986,Chubukov_1991,Rastelli_1996} The
magnetic phase diagrams of such 2D magnets strongly depend on the spin value of
magnetic ions.

CuCrO$_2$ is an example of quasi-two-dimensional antiferromagnet ($S=3/2$) with triangular lattice structure. Below $T_N\approx 24$~K CuCrO$_2$ exhibits spiral ordering to incommensurate spiral magnetic structure with a small deviation from regular 120$^\circ$ structure. The transition to the magnetically ordered state is accompanied by a small distortion of triangular lattice. We present a NMR study of low temperature magnetic structure of CuCrO$_2$ in the fields up to 15.5~T. These fields are small in comparison with exchange interactions within the triangular plane ($\mu_0H_{sat}\approx 280$~T). Thus, we can expect that in our experiments the exchange structure within individual plane will not be distorted significantly and the field evolution of NMR spectra in our experiments will be due to spin plane reorientation or change of interplane ordering. The microscopic properties of magnetic phases of this magnet are especially interesting because this material is multiferroic.~\cite{Seki_2008,Kimura_PRB_2008,Kimura_PRB_2009} The possibility to modify electric and magnetic domains with electric and magnetic fields make CuCrO$_2$ attractive for experimental study of the magnetoelectric coupling in this class of materials.

\section{Crystal and magnetic structure}

\begin{figure}[b!]
\includegraphics[width=.55\columnwidth,angle=0,clip]{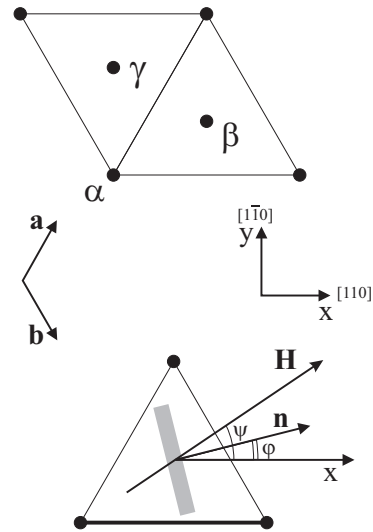}
\caption{(color online) Top: Crystal structure of CuCrO$_2$ projected on the $ab$-plane. The three layers, $\alpha\beta\gamma$, are the positions of Cr$^{3+}$ ions. Bottom: Reference angles $\psi$ and $\varphi$ as defined in the text; the gray bar corresponds to the projection of spin plane. The incommensurate wavevector $\vect{q}_{ic}$ is collinear with the base of the triangle (thick line).}
\label{fig1}
\end{figure}

The structure CuCrO$_2$ consists of magnetic Cr$^{3+}$ (3$d^3$, $S=3/2$), nonmagnetic Cu$^+$, and O$^{2-}$ triangular lattice planes (TLPs), which are stacked along $c$-axis in the sequence Cr-O-Cu-O-Cr (space group $R\bar{3}m$, $a=2.98$~\AA{}, $c =17.11$~\AA{} at room temperature~\cite{Poienar_2009}). The layer stacking sequences are $\alpha\gamma\beta$, $\beta\alpha\gamma$, and $\beta\beta\alpha\alpha\gamma\gamma$ for Cr, Cu and O ions, respectively. The crystal structure of CuCrO$_2$ projected on the $ab$-plane is shown on top portion of Fig.~1. The distances between the nearest planes denoted by different letters for copper and chromium ions and the pairs of planes for oxygen ions are $c/3$, whereas the distance between the nearest oxygen planes denoted by the same letters is $(1/3-0.22)c$ ~(Ref.~[\onlinecite{Poienar_2009}]). No structural phase transition has been reported at temperatures higher than N\'{e}el ordering temperature ($T > T_N\approx 24$~K). In the magnetically ordered state the triangular lattice is distorted, so that one side of the triangle becomes slightly smaller than two other sides: $\Delta a / a \simeq 10^{-4}$~(Ref.~[\onlinecite{Kimura_JPSJ_2009}]).

The magnetic structure of CuCrO$_2$ has been intensively investigated by neutron diffraction experiments.~\cite{Poienar_2009, Kadowaki_1990, Soda_2009, Soda_2010, Frontzek_2012} It was found that the magnetic ordering in CuCrO$_2$ occurs in two stages.~\cite{Frontzek_2012, Aktas_2013} At the higher transition temperature $T_{N1}=24.2$~K two dimensional (2D) ordered state within $ab$-planes sets in, whereas below $T_{N2}=23.6$~K three dimensional (3D) magnetic order with incommensurate propagation vector $\vect{q}_{ic}= (0.329, 0.329, 0)$ along the distorted side of TLPs~\cite{Kimura_JPSJ_2009} is established. The magnetic moments of Cr$^{3+}$ ions can be described by the expression
\begin{eqnarray}
\vect{M}_i=M_1\vect{e}_1\cos(\vect{q}_{ic}\cdot\vect{r}_i+\theta)+M_2\vect{e}_2\sin(\vect{q}_{ic}\cdot\vect{r}_i+\theta),
\label{eqn:spiral}
\end{eqnarray}
where $\vect{e}_1$ and $\vect{e}_2$ are two perpendicular unit vectors determining the spin plane orientation with the normal vector $\vect{n}=\vect{e}_1 \times \vect{e}_2$, $\vect{r}_i$ is the vector to the $i$-th magnetic ion and $\theta$ is an arbitrary phase. The spin plane orientation and the propagation vector of the magnetic structure are schematically shown in the bottom of Fig.~1. For zero magnetic field $\vect{e_1}$ is parallel to $[001]$ with $M_1 = 2.8(2)~\mu_B$, while $\vect{e_2}$ is parallel to $[1\bar{1}0]$ with $M_2 = 2.2(2)~\mu_B$ (Ref.~[\onlinecite{Frontzek_2012}]). The pitch angle between the neighboring Cr moments corresponding to the observed value of $\vect{q}_{ic}$ along the distorted side of TLP is equal to 118.5$^\circ$ which is very near to 120$^\circ$ expected for regular TLP structure.

Owing to the crystallographic symmetry in the ordered phase we can expect {\it six} magnetic domains at $T < T_N$. The propagation vector of each domain can be directed along one side of the triangle and can be positive or negative. As reported in Refs.~[\onlinecite{Soda_2010, Svistov_2013}], the distribution of the domains is strongly affected by the cooling history of the sample.

Inelastic neutron scattering data~\cite{Poienar_2010} has shown that CuCrO$_2$ can be considered as a quasi 2D magnet. The spiral magnetic structure is defined by the strong exchange interaction between the nearest Cr ions within the TLPs with exchange constant $J_{ab}=2.3$~meV. The inter-planar interaction is at least one order of magnitude weaker than the in-plane interaction.

Results of the magnetization, ESR and electric polarization experiments~\cite{Svistov_2013, Kimura_PRB_2009} has been discussed within the framework of the planar spiral spin structure at fields studied experimentally: $\mu_0H < 14$~T~$\ll \mu_0H_{sat}$ ($\mu_0H_{sat}\approx 280$~T). The orientation of the spin plane is defined by the biaxial crystal anisotropy. One {\it hard} axis for the normal vector $\vect{n}$ is parallel to the $c$ direction and the second axis is perpendicular to the direction of the distorted side of the triangle. The anisotropy along $c$ direction dominates with anisotropy constant approximately hundred times larger than that within $ab$-plane resulting from the distortions of the triangle structure. A magnetic phase transition was observed for the field applied perpendicular to one side of the triangle ($\vect{H}\parallel [1\bar{1}0]$) at $\mu_0H_c = 5.3$~T, which was consistently described~\cite{Soda_2010, Svistov_2013, Kimura_PRB_2009} by the reorientation of the spin plane from $(110)$ ($\vect{n}\perp\vect{H}$) to $(1\bar{1}0)$ $(\vect{n}\parallel\vect{H})$. This spin reorientation happens due to weak susceptibility anisotropy of the spin structure $\chi_{\parallel}\approx 1.05\chi_{\perp}$.

\section{Sample preparation and experimental details}

A single crystal of CuCrO$_2$ was grown by the flux method following~Ref.~[\onlinecite{Frontzek_2012}]. The crystal structure was confirmed by single crystal room-temperature x-ray spectroscopy. The magnetic susceptibility ($M(T)/H$) was measured at $\mu_0H=0.5$~T in the temperature range from 2 to 300~K using SQUID magnetometer. The obtained susceptibility curve $\chi_c$ was similar to the data from~Refs.~[\onlinecite{Okuda_2005, Kimura_PRB_2008}]. The N\'{e}el temperature $T_N\approx 24$~K and the Curie-Weiss temperature $\theta_{CW}=-204$~K obtained from the fitting of $M(T)$ at temperatures $150<T<300$~K are in agreement with the values given in~Ref.~[\onlinecite{Kimura_PRB_2008}].

NMR experiments were carried out using a home-built NMR spectrometer. Measurements were taken on a 17.5~T Cryomagnetics field-sweepable NMR magnet at the National High Magnetic Field Laboratory. $^{63,65}$Cu nuclei (nuclear spins $I=3/2$, gyromagnetic ratios $\gamma^{63}/2\pi=11.285$~MHz/T, $\gamma^{65}/2\pi=12.089$~MHz/T) were probed using pulsed NMR technique. In the figures that follow, the spectra shown by solid lines were obtained by summing fast Fourier transforms (FFT), while the spectra shown by circles were obtained by integrating the averaged spin-echo signals as the field was swept through the resonance line. NMR spin echoes were obtained using $1.5~\mu$s$~-~\tau_D~-~3~\mu$s ($\vect{H}\parallel\vect{c}$), $1.8~\mu$s$~-~\tau_D~-~3.6~\mu$s ($\vect{H}\perp\vect{c}$, $\mu_0H\sim 4.5$~T), $2.3~\mu$s$~-~\tau_D~-~4.6~\mu$s ($\vect{H}\perp\vect{c}$, $\mu_0H\sim 11.6$~T) pulse sequences, where the times between pulses $\tau_D$ were 15, 20, 15~$\mu$s, respectively. Measurements were carried out in the temperature range $4.2\leq T \leq 40$~K stabilized with a precision better than 0.03~K. The experimental setup allowed rotating the sample inside the excitation coil with respect to the static field $\vect{H}\perp\vect{c}$ during the experiment.

\section{Experimental results}

The $^{63,65}$Cu NMR spectra for the paramagnetic (Figs.~2a,3a) and ordered (Figs.~2b,3b) states for $\vect{H}\perp\vect{c}$ and $\vect{H}\parallel\vect{c}$ consist of two sets of lines, corresponding to $^{63}$Cu and $^{65}$Cu isotopes. Each set consists of three lines; one of them corresponds to the central line ($m_I = -1/2\leftrightarrow +1/2$) and two quadrupole satellites corresponding to ($\pm 3/2\leftrightarrow \pm 1/2$) transitions.

For the paramagnetic state the spectral shape was found to be independent of the magnetic field orientation within $ab$-plane. In contrast, the shapes of the spectra at temperatures below $T_N$ are strongly dependent on the field direction and cooling history. The spectra were studied under two cooling conditions: zero field cooling (ZFC) and field cooling (FC).

In the first set of the experiments the static field was oriented within $ab$-plane (Figs.~4-9). The static magnetic field during the FCs was directed parallel ($\vect{H}\parallel [110]$) or perpendicular ($\vect{H}\parallel [1\bar{1}0]$) to one side of the triangular structure. For all FC data, the sample was cooled in a field of 16.9~T from 40 to 4.2~K (or 5~K) with characteristic time $t_{cooling}\approx 40$~min, and then the measurements were performed. The sample was rotated about the $c$-axis. The direction of the external field given on the figures is measured with respect to the $[110]$ direction of the sample. Since the data obtained for the full rotation of the sample reveal a $180^\circ$ symmetry only data from $0^\circ$ to $150^\circ$ are shown for clarity.

\begin{figure}
\includegraphics[width=0.9\columnwidth,angle=0,clip]{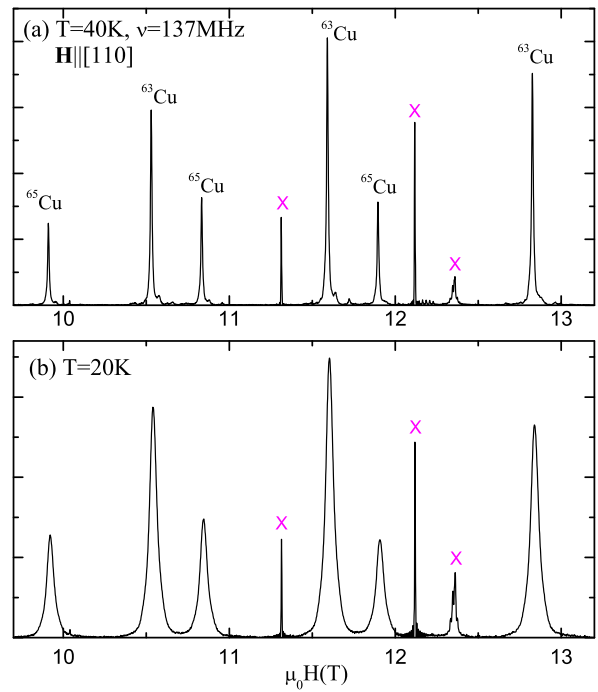}
\caption{(color online) $^{63,65}$Cu NMR spectra of the CuCrO$_2$ single crystal in the paramagnetic state (a) and in the ordered state (b) at the external magnetic field directed perpendicular to $c$ axis, $\vect{H}\parallel[110]$. The two sets of lines correspond to the signals from quadrupolar split $^{63}$Cu and $^{65}$Cu nuclei (see text). The peaks marked with crosses are spurious $^{63,65}$Cu and $^{27}$Al NMR signals from the probe.}
\label{fig2}
\end{figure}

\begin{figure}
\includegraphics[width=0.9\columnwidth,angle=0,clip]{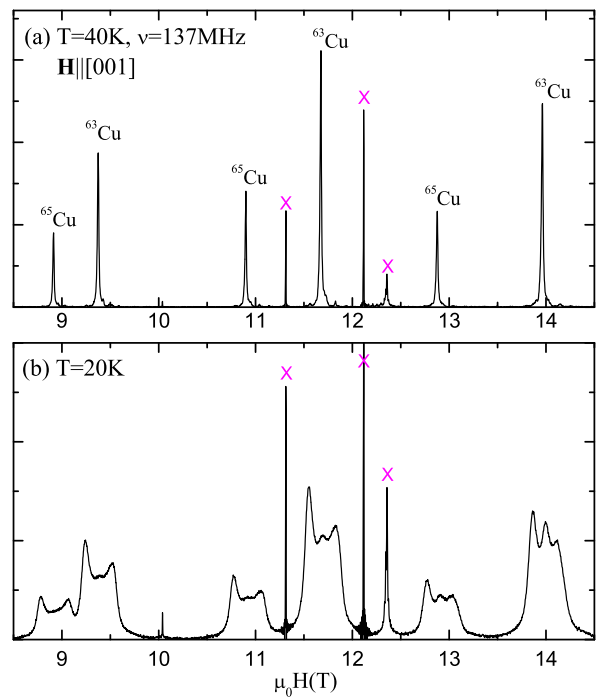}
\caption{(color online) Similar spectra as in Fig.~2 except with field applied parallel to $c$ axis, $\vect{H}\parallel[001]$.}
\label{fig3}
\end{figure}

The angular dependences within $ab$-plane were measured at the frequencies 55.3~MHz and 137~MHz. The lower frequency is chosen such that the central line of $^{63}$Cu NMR is situated at fields near 4.5~T, i.e. below reorientation field $\mu_0H_c=5.3$~T, whereas the higher frequency will place the resonance above $\mu_0H_c$.

The NMR spectra were measured at two temperatures $T_{high}=20$~K (Figs.~4,8) and $T_{low}=4.2$~K (Figs.~5,6,7) and $T_{low}=5$~K (Fig.~9). We chose these two temperature sets, since the domain walls in CuCrO$_2$ are mobile at high temperatures, whereas at low temperatures the walls are pinned.~\cite{Svistov_2013}

In the second set of the experiments the static field was oriented parallel to the $c$ direction. Representative ZFC spectra at 20~K measured at different fields are shown in Fig.~10.

\section{Discussion}

We shall discuss the results of the NMR experiments in the framework of planar spiral spin structure (Eq.~(\ref{eqn:spiral})) proposed from neutron diffraction experiments~\cite{Frontzek_2012} and carried out at $H=0$. Spin plane orientation of such structure is defined by weak relativistic interactions with the external field and crystal environment. Following~Ref.~[\onlinecite{Svistov_2013}], the anisotropic part of magnetic energy of CuCrO$_2$ can be written as:

\begin{eqnarray}
U=-\frac{\chi_{\parallel}-\chi_{\perp}}{2}(\vect{n}\vect{H})^2+\frac{A}{2}n_{z}^2+\frac{B}{2}n_{y}^2,
\label{eqn:energy}
\end{eqnarray}
where $A > B > 0$. For the arbitrary field direction within $ab$-plane the vector $\vect{n}$ will monotonously rotate from $\vect{n}\parallel[110]$ to $\vect{n}\parallel\vect{H}$. This can be defined by minimization of Eq.~(\ref{eqn:energy}), which can be rewritten as:

\begin{eqnarray}
U=-\frac{\Delta\chi}{2}H^2(\cos^2(\psi - \varphi)-\Bigr(\frac{H_{cy}}{H}\Bigl)^2\sin^2\varphi)+\frac{A}{2}n_{z}^2.
\label{eqn:energy_psi}
\end{eqnarray}
Here the angles $\psi$ and $\varphi$ define the directions of the vectors $\vect{H}$ and $\vect{n}$, respectively, as depicted in Fig.~1. Using value of reorientation field $\mu_0H_{cy}=5.3$~T we obtain the expected orientations of the spin planes relative to the field directions of our NMR experiments. For the field directed along ``thick'' side of the triangle within $ab$-plane (i.e. $\psi=0^\circ$) $\varphi=0^\circ$ at all fields. For the field directed along ``thin'' side of the triangle (i.e. $\psi=60^\circ$) the expected angles $\varphi$ for a given field are: $\varphi(\mu_0H=4.5$~T$)=22.15^\circ$ and $\varphi(\mu_0H=11.6$~T$)=54.3^\circ$. For the field direction perpendicular to ``thick'' side of triangle (i.e. $\psi=90^\circ$) below spin-flop $\varphi(\mu_0H < \mu_0H_{cy}=5.3$~T$)=0^\circ$ and above spin-flop $\varphi(\mu_0H > \mu_0H_{cy}=5.3$~T$)=90^\circ$. For the field direction perpendicular to ``thin'' side of triangle (i.e. $\psi=30^\circ$) the orientation of spin plane is defined by the angles: $\varphi(\mu_0H=4.5$~T$)=12.3^\circ$ and $\varphi(\mu_0H=11.6$~T$)=25.4^\circ$.

\begin{figure}
\includegraphics[width=1\columnwidth,angle=0,clip]{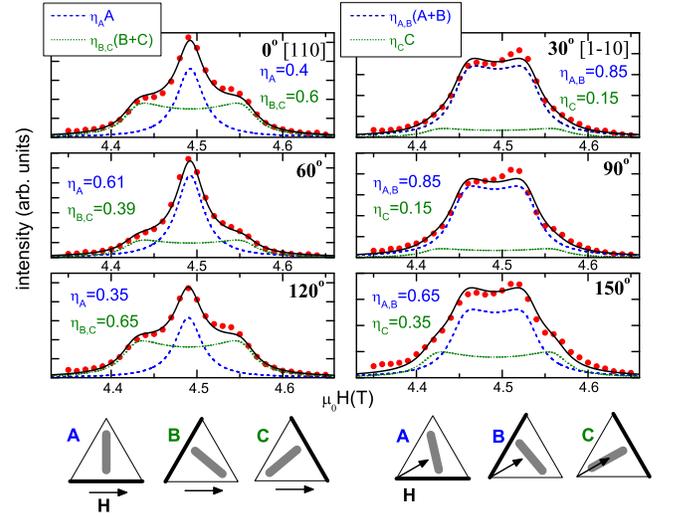}
\caption{(color online) $^{63}$Cu NMR spectra ($m_I = -1/2\leftrightarrow +1/2$) measured at different angles between $\vect{H}$ applied within $ab$-plane and $[110]$ direction of the sample (red solid circles). ZFC to $T=20$~K, $\nu=55.3$~MHz. Black solid lines on the figure are calculated spectra in the model of the magnetic structure (Eq.~(\ref{eqn:spiral})) and orientations of spin planes (gray bars) shown in the bottom of the figure. $A$, $B$ and $C$ accord to three possible alignment of propagation vector of magnetic structure ($\vect{q}_{ic}$ is collinear to the triangle side marked thick). $\eta_A$, $\eta_B$, $\eta_C$ - the relative weights of the NMR signals from $A$, $B$, $C$ domains, which were used for the best coincidence with experiment.}
\label{fig4}
\end{figure}

\begin{figure}
\includegraphics[width=1\columnwidth,angle=0,clip]{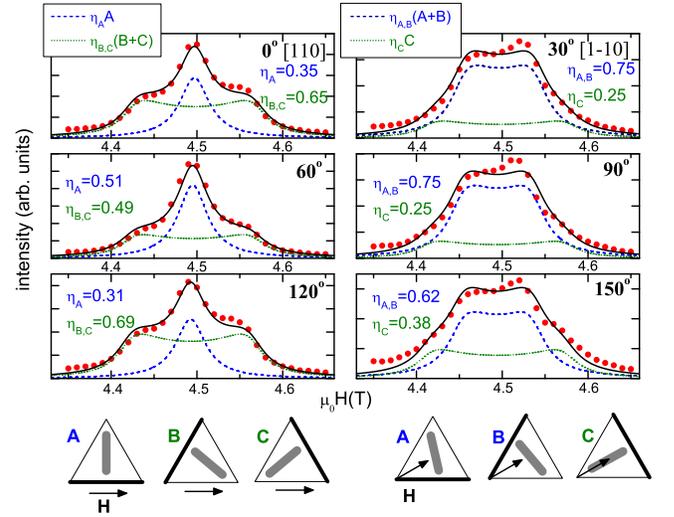}
\caption{(color online) Similar data to Fig.~4 except ZFC to $T=4.2$~K.}
\label{fig5}
\end{figure}

For $\vect{H}\parallel\vect{c}$ the field of spin reorientation transition $H_{cz}$ is expected much larger, than the fields in our experiments.~\cite{Svistov_2013} Therefore the spin plane orientation for $\vect{H}\parallel\vect{c}$ we expect the same as at $H=0$ ($\vect{n}\parallel [110]$).

Analyzing the NMR spectral shapes we found that they can be well described
within the model of spiral spin structure given by~Eq.~(\ref{eqn:spiral}) with
the incommensurate propagation vector directed along one side of the triangle
$q_{ic}=0.329$. Generally, the local magnetic field at Cu sites is the sum of
the long range dipole field $\vect{H}_{dip}$ and the transferred hyperfine
contact field produced by the nearest Cr$^{3+}$ moments. The $^{63,65}$Cu NMR
of CuCrO$_2$ in paramagnetic state was studied
in~Ref.~[\onlinecite{Verkhovskii_2011}], where it was shown that the effective
field at the copper site is proportional to the chromium moment with a
hyperfine field of 3.3~T/$\mu_B$. This value does not depend on the direction
of the chromium moment. The computed dipolar fields on the chromium nuclei in
paramagnetic phase are anisotropic and essentially smaller, than experimentally
observed, and equal to 0.17~T/$\mu_B$ and -0.08~T/$\mu_B$ for
$\vect{H}\parallel[001]$ and $\vect{H}\parallel[1\bar{1}0]$, respectively. From
these data we presumed that the effective field measured in paramagnetic
phase~\cite{Verkhovskii_2011} is mostly defined by the contact fields from six
nearest chromium magnetic ions. Although the contact field created by an
individual neighbor chromium moment (3.3/6~T/$\mu_B=0.55$~T/$\mu_B$) is much
larger than the dipole field, dipolar and contact hyperfine contributions prove
to be comparable in the ordered state. This is due to a strong compensation of
the contact fields from the six nearest chromium moments in the ordered state.
We took into account both contributions in our calculations.
The dipolar fields on the copper nuclei were computed by numerically summing
contributions from nearest neighbor Cr moments in the sphere of radius
$20$~\AA{}, further accounting for the moments farther away gives no noticeable effect.
The shape of the individual NMR line was taken Lorentzian for the fits.

The best fit with experiment $\vect{H}\perp\vect{c}$ was obtained with the value of Cr$^{3+}$ magnetic moments $M_1=M_2=0.91(5)~\mu_B$ at low fields $\mu_0H\sim 4.5$~T (Figs.~4,5,6,7) and $M_1=M_2=1.15(9)~\mu_B$ at high fields $\mu_0H\sim 11.6$~T (Figs.~8,9). Each individual linewidth is $\delta=20(5)$~mT for all fitted NMR spectra. This value is consistent with the linewidth measured in the paramagnetic phase.

Since the copper nuclei in CuCrO$_2$ are situated at a position of high symmetry, the NMR spectra from the magnetic domains with opposite directions of incommensurate vectors $+\vect{q}_{ic}$ and $-\vect{q}_{ic}$ are identical. In the analysis that follow we shall assign magnetic domains as letters $A$, $B$ and $C$, having in mind that each letter refers to two magnetic domains indistinguishable by NMR method.

The ZFC NMR spectra measured at frequency 55.3~MHz and temperatures 20~K and 4.2~K are shown on Figs.~4 and 5. For the field directed parallel to one side of the triangle ($\psi=0^\circ$, $60^\circ$, $120^\circ$) we expect that the resonance conditions of two domains ($B$, $C$) will be equivalent. A sketch of the expected spin plane orientations within $A$~($\psi=0^\circ$, $\varphi=0^\circ$), $B$~($\psi=60^\circ$, $\varphi=22.15^\circ$), $C$~($\psi=120^\circ$, $\varphi=157.85^\circ$) domains are shown at the bottom left of each figure. The dashed and dotted lines show the computed spectra for domains $A$ and $B+C$, respectively. The resulting spectrum has been obtained by summing the spectra from the three domains with relative weights $\eta_A$, $\eta_B$, $\eta_C$ and is shown as solid lines in the figures.

\begin{figure}
\includegraphics[width=1\columnwidth,angle=0,clip]{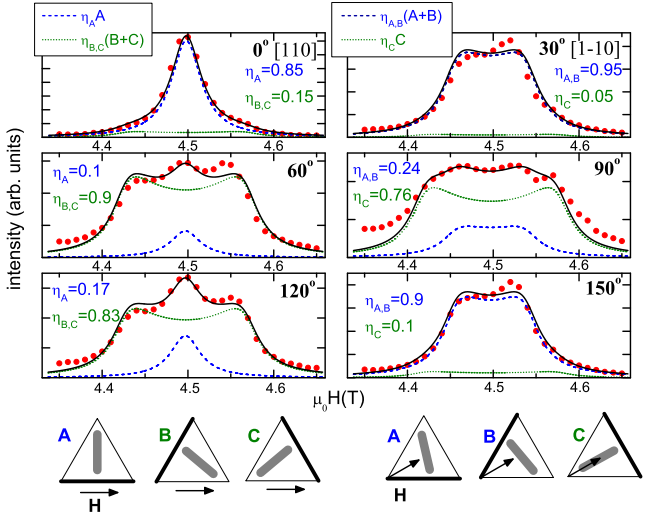}
\caption{(color online) Similar data to Fig.~4 except FC $\vect{H}$(16.9~T)$\parallel[110]$ ($\psi=0^\circ$) to $T=4.2$~K. Background at some panels are due to overlapping with $^{65}$Cu NMR lines.}
\label{fig6}
\end{figure}

\begin{figure}
\includegraphics[width=1\columnwidth,angle=0,clip]{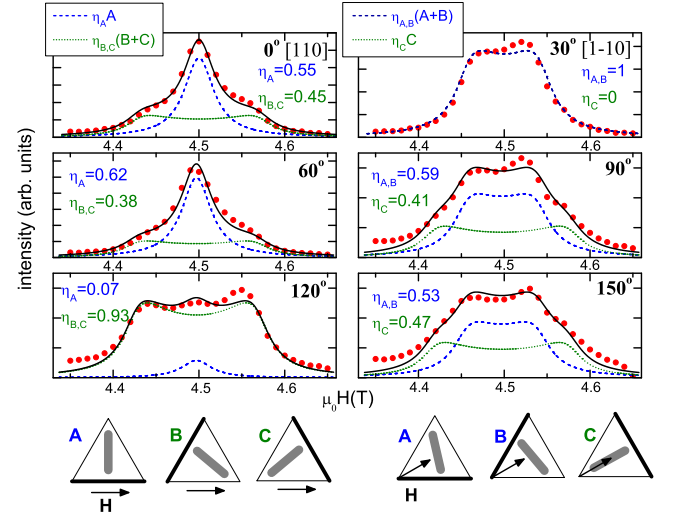}
\caption{(color online) Similar data to Fig.~4 except FC $\vect{H}$(16.9~T)$\parallel[1\bar{1}0]$ ($\psi=30^\circ$) to $T=4.2$~K. Background at some panels are due to overlapping with $^{65}$Cu NMR lines.}
\label{fig7}
\end{figure}

If the relative sizes of the domains do not change during the rotation of the field we expect that the sum of the relative fractions of domain $A$ $\eta_A$ measured at three unique orientations $\psi=0^\circ$, $60^\circ$, $120^\circ$ i.e., $\eta_A(0^\circ)+\eta_A(60^\circ)+\eta_A(120^\circ)$, will be equal to unity. The same is true for the sum $\eta_C(30^\circ)+\eta_C(90^\circ)+\eta_C(150^\circ)$. However, the experimental values of the sums for $T=20$~K (Fig.~4) are $\eta_A(0^\circ)+\eta_A(60^\circ)+\eta_A(120^\circ)=1.36$ and $\eta_C(30^\circ)+\eta_C(90^\circ)+\eta_C(150^\circ)=0.65$. This deviation from unity shows that the domain sizes of the sample change with field rotation. These observations show that the size of the energetically favorable domains grows in expense of unfavorable domains. To our knowledge, this phenomenon is directly observed the first time through the NMR technique.

For measurements at $T=4.2$~K (Fig.~5) the sums are more close to unity: $\eta_A(0^\circ)+\eta_A(60^\circ)+\eta_A(120^\circ)=1.17$; $\eta_C(30^\circ)+\eta_C(90^\circ)+\eta_C(150^\circ)=0.88$. This implies that the mobility of domain walls increases with the temperature. We emphasize that the conversion of the domain structure implies changes not only in the spin plane orientation, but also in the direction of the wave vector $\vect{q}_{ic}$.

The field cooling of the sample enables us to prepare the sample with the energetically preferable domains. If the field during the cooling process was directed along one side of the triangle ($\psi=0^\circ$, Fig.~6) domain $A$ is preferable. In this case, field cooling the sample results in $\sim 85$~\% domain $A$ and $\sim 15$~\% $B$ and $C$. If the cooling field is directed perpendicular to one side of the triangle ($\psi=30^\circ$, Fig.~7) the domains $B$ and $C$ are preferable. In such a case, the NMR signal from unfavorable domain will be negligibly small and the two other domains will have nearly the same sizes. Interestingly, the relative part of the sample, where the direction of $\vect{q}_{ic}$ changes with field rotation at $T=4.2$~K has nearly the same intensity as for the ZFC procedure. The parameters defining the domain conversion for FC samples are $\eta_A(0^\circ)+\eta_A(60^\circ)+\eta_A(120^\circ)=1.12$ and $\eta_C(30^\circ)+\eta_C(90^\circ)+\eta_C(150^\circ)=0.91$ for the cooling field direction $\psi=0^\circ$ (Fig.~6). For the cooling field direction $\psi=30^\circ$ (Fig.~7) these parameters are 1.24 and 0.88, respectively. Thus, the domain conversion during the rotation of the static field $\mu_0H\approx4.5$~T at $T=4.2$~K takes place within 4-8~\% of the sample.

Such domain conversion is more definitely observed at higher fields $\mu_0H\approx11.6$~T $>\mu_0H_c=5.3$~T and higher temperature $T=20$~K (see Fig.~8). For the fields directed along one side of the triangular structure, the NMR spectra can be solely identified with a single energetically preferable domain with $\vect{q}_{ic}$ parallel to the applied field ($\psi=0^\circ$, $60^\circ$, $120^\circ$, spectra on the left panels of Fig.~8). This means, that at $T=20$~K, the field ($\sim11.6$~T) rotates not only the spin plane of magnetic structure, but also rebuilds the domain structure, so that only the energetically preferable domains are established in the sample, namely the domains with $\vect{q}_{ic}\parallel\vect{H}$. For the fields directed perpendicular to one side of the triangle ($\psi=30^\circ$, $90^\circ$, $150^\circ$, spectra on the right panels of Fig.~8) domains $A$ and $B$ are more energetically preferable. The ZFC spectra observed at high fields ($\sim11.6$~T) and low temperature (5~K) (Fig.~9) are qualitatively similar to ZFC spectra measured at low fields ($\sim4.5$~T). Only the sums defining the domain conversion of the sample, $\eta_A(0^\circ)+\eta_A(60^\circ)+\eta_A(120^\circ)=1.27$ and $\eta_C(30^\circ)+\eta_C(90^\circ)+\eta_C(150^\circ)=0.5$, are larger than those for the low field case (1.17 and 0.88, respectively).

\begin{figure}
\includegraphics[width=1\columnwidth,angle=0,clip]{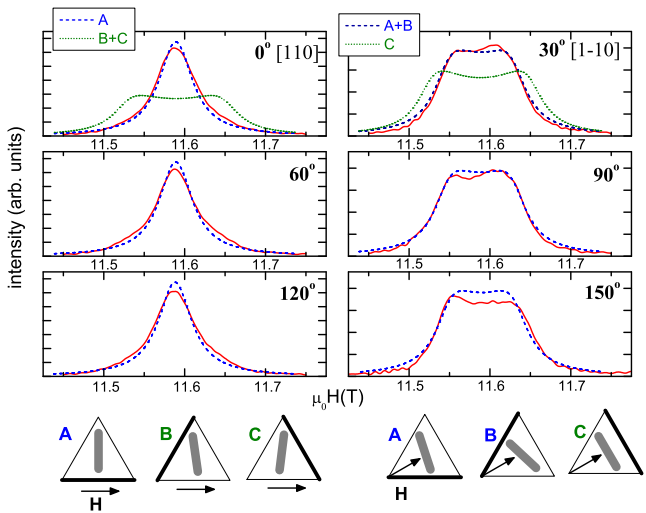}
\caption{(color online) Similar data to Fig.~4 except ZFC to $T=20$~K, $\nu=137$~MHz.}
\label{fig8}
\end{figure}

\begin{figure}
\includegraphics[width=1\columnwidth,angle=0,clip]{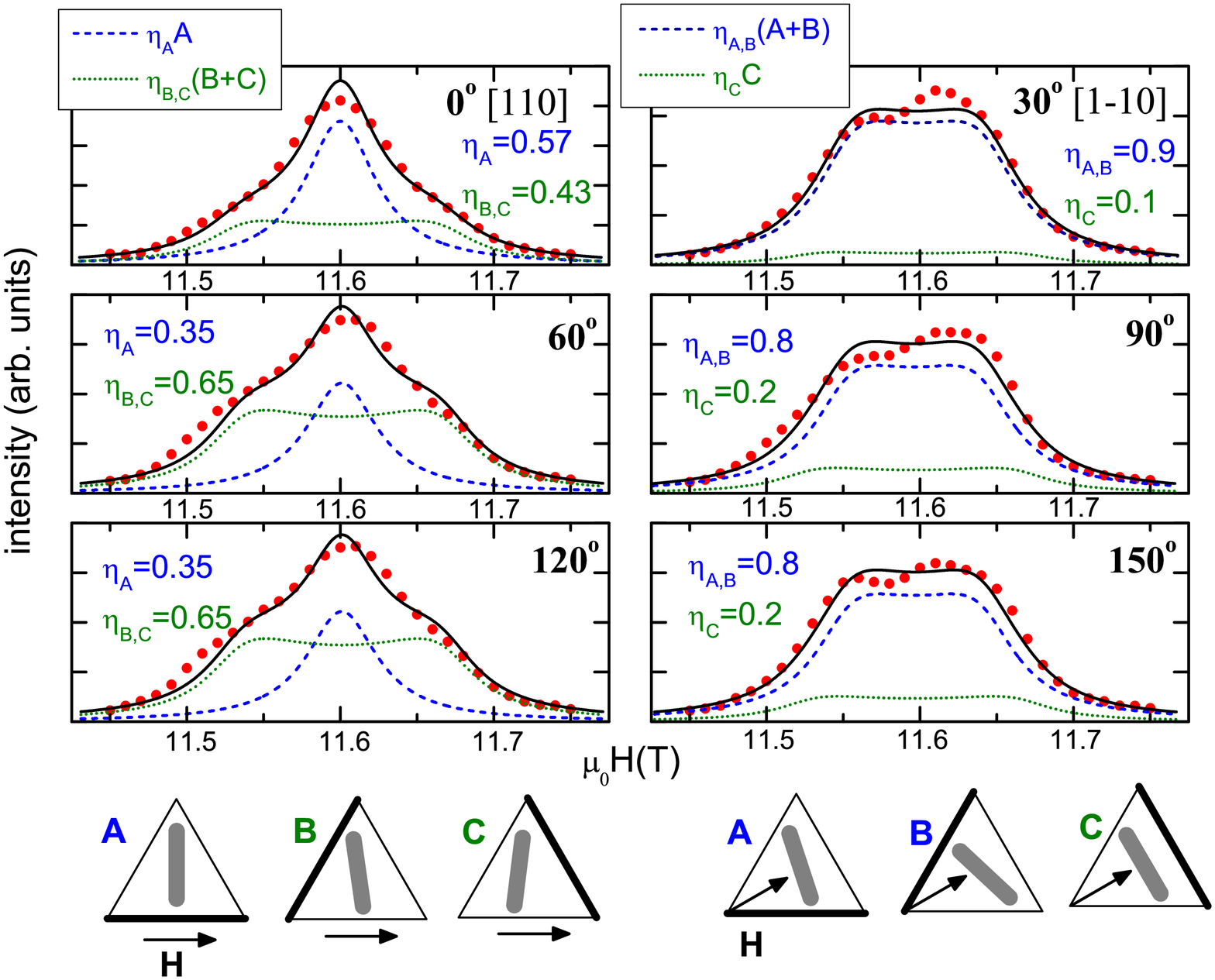}
\caption{(color online) Similar data to Fig.~4 except ZFC to $T=5$~K, $\nu=137$~MHz.}
\label{fig9}
\end{figure}

The NMR spectra measured at field along the $c$-axis ({\it hard} axis for $\vect{n}$ vector of the structure) is 2.5 times broader, than those with fields aligned within the $ab$-plane, Fig.~10. The shape of the spectra depends on the field value. At low field range ($\mu_0H<\sim10$~T) the shape has two horns similar to those observed for fields oriented within $ab$-plane. For higher fields, an additional third peak appears in the middle of the spectra.

The low field spectra can be satisfactorily described by the model of incommensurate spiral spin structure similar to the structure proposed for zero field (Eq.~(\ref{eqn:spiral})). The best fit at $T=20$~K and $\mu_0H\approx9$~T is obtained with $M_1=M_2=2.1~\mu_B$, $\delta=20$~mT (Fig.~10, blue dotted line).

\begin{figure}
\includegraphics[width=0.95\columnwidth,angle=0,clip]{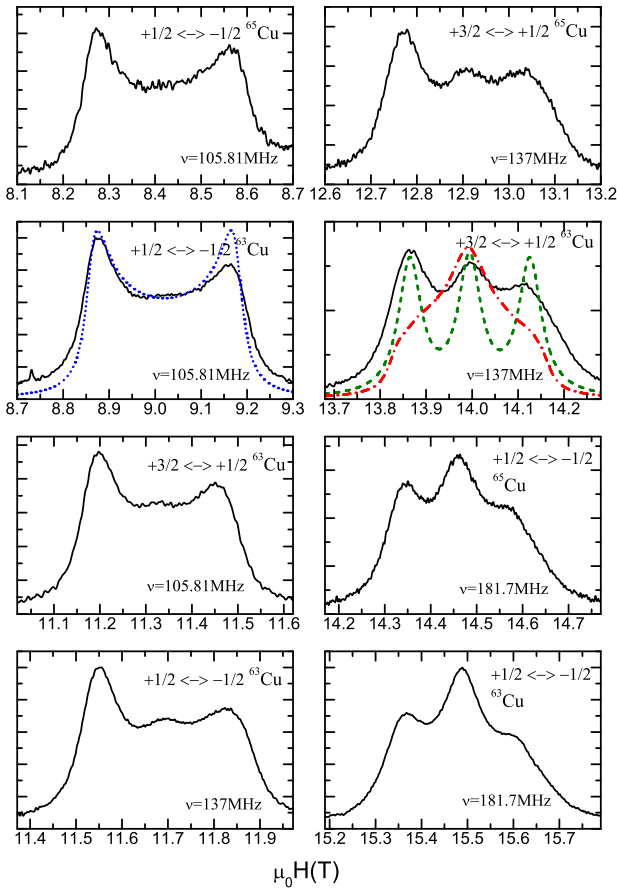}
\caption{(color online) Representative $^{63,65}$Cu NMR spectra at different field (black solid lines) under ZFC condition, $T=20$~K. Shown are data taken with each field swept over a narrow range at fixed frequencies. The $x$-axis was adjusted so that each panel covers the same field range. The blue dotted line is calculated spectrum using (Eq.~(\ref{eqn:spiral})), red dot-dashed and green dashed lines are calculated spectra corresponding to incommensurate spiral magnetic structure with disorder in the $c$ direction and commensurate spiral magnetic structure. See text for details.}
\label{fig10}
\end{figure}

The change of the NMR spectral shape at higher fields indicates the field evolution of the magnetic structure. We suggest two possible models of the high field magnetic structure. The red dash-dotted line in Fig.~10 shows the result of computed NMR spectra using Eq.~(\ref{eqn:spiral}) with random phases $\theta$ for structures within different triangular planes with $M_1=M_2=2.2~\mu_B$, $\delta=20$~mT. Such model accounts for the incommensurate spiral magnetic structure within every $ab$-plane with disorder in the $c$ direction. In this case the experimental spectra can be described by the sum of the spectra from the parts of the sample with order and disorder along $c$ direction.

Another model for the three horn spectra is the commensurate magnetic structure with pitch angle close to $120^\circ$, in which one moment looks along $[00\bar{1}]$ direction, i.e. opposite to the applied field $\vect{H}$. Computed NMR spectrum for this case with $M_1=M_2=2.7~\mu_B$, $\delta=30$~mT is given in the Fig.~10, shown as green dashed line. A transition of the spiral structure from incommensurate to commensurate in large enough fields has been observed in other compounds with triangular structure.~\cite{Prozorova_1993, Prozorova_2004, Park_2007}

For both models the values of $M_1$, $M_2$ are closer to the maximum value expected for chromium magnetic moment $g\mu_BS$, compare to those values for perpendicular orientation. It is probable that the static field applied along $c$ direction suppresses the fluctuating part of the magnetic moments of Cr ions which could explain its large value when $\vect{H}\parallel\vect{c}$ compared to its value when $\vect{H}\perp\vect{c}$.

The observed ``third peak'' peculiarity possibly corresponds to a phase transition also seen by recent electric polarization experiments.~\cite{Zapf_2014}

\section{Conclusions}

$^{63,65}$Cu NMR spectra measured at $\vect{H}\perp[001]$ can be well described by the planar spiral magnetic structure with the oscillating components of the moments approximately 2.5 times smaller than obtained from neutron diffraction experiments.~\cite{Poienar_2009, Frontzek_2012} Rotation of the sample in a magnetic field results in the reorientation of the spin plane accompanied by the reorientation of the incommensurate wave vector of the structure. This wave vector follows the direction of magnetic field at high enough temperature and fields, whereas at low temperatures or low fields the propagation vector is defined by the cooling history of the sample. The results are consistent with previous results of electric polarization and ESR studies.~\cite{Svistov_2013, Kimura_PRB_2009}

The NMR study of the magnetic structure at $\vect{H}\parallel[001]$ shows that the low field magnetic structure is consistent with the structure proposed by neutron diffraction experiments. In fields higher than 10~T the magnetic structure is modified. The results can be described by the loss of long range ordering in the $c$ direction, or by the transition from incommensurate to commensurate structure. This observation opens interesting possibilities for future experimental investigations.

\acknowledgements We thank V.I. Marchenko and V.L. Matukhin for stimulating
discussions. Yu.A.S. is grateful to the Institute of International Education,
Fulbright Program, for financial support (Grant 68435029). H.D.Z. thanks for
the support from  NSF-DMR through award DMR-1350002. This work was supported by
Russian Foundation for Basic Research, Program of Russian Scientific Schools
(Grant 13-02-00637). Work at the National High Magnetic Field Laboratory is
supported by the NSF Cooperative Agreement No. DMR-0654118, and by the State of
Florida.

\end{document}